\documentclass[11pt,a4paper]{article}
\usepackage{authblk}
\usepackage[hyperref]{emnlp-ijcnlp-2019}
\usepackage{times}
\usepackage{latexsym}
\usepackage{graphicx}
\usepackage{amsmath,amssymb,caption,array,bm,hyperref}
\usepackage{algorithm,algorithmic,color,multirow,float,subcaption}

\usepackage{url}

\newcommand{\vw}{\textbf{w}}
\newcommand{\bw}{\textbf{W}}

\aclfinalcopy 

\title{Shallow Domain Adaptive Embeddings for Sentiment Analysis}

\author[]{Prathusha K Sarma}
\author[]{Yingyu Liang}
\author[]{William A Sethares}

\affil[]{University of Wisconsin-Madison \\
         {\tt \{kameswarasar,sethares\}@wisc.edu},\\
         \tt yliang@cs.wisc.edu}

\date{}

\begin{document}
\maketitle
\begin{abstract}
This paper proposes a way to improve the performance of existing algorithms for
text classification in domains with strong language semantics.
We propose a domain adaptation layer learns weights to combine a generic and a domain
specific (DS) word embedding into a domain adapted (DA) embedding. The DA word embeddings
are then used as inputs to a generic encoder + classifier framework to perform a downstream task such
as classification. This adaptation layer is particularly suited to datasets that are modest in size, and which are, therefore, not ideal candidates for (re)training a deep neural network architecture. Results on binary and multi-class classification tasks using popular encoder architectures, including current state-of-the-art methods (with and without the shallow adaptation layer) show the effectiveness of the proposed approach.
\end{abstract}

\section{Introduction}

Domain Adaptation (DA) algorithms are becoming increasingly relevant in addressing issues related to i) lack of availability of training data in domains of interest and in ii) exploiting domain idiosyncrasies to improve performance of out-of-domain algorithms. While some state of the art DA algorithms focus on improving on a downstream task, such as classification in bilingual or crosslingual framework~\cite{P18-1075}~\cite{N18-1050}, others~\cite{D18-1041},~\cite{D18-1230} focus on tackling DA at the word level for various downstream applications such as classification, tagging etc.

Particularly, work by~\cite{P18-1228} and~\cite{P18-2007} address the issue of context-based domain adaptation within a single language. The authors of both works argue that word context causes significant changes, especially in document level sentiment associations. The idea is illustrated by providing examples of words such as `kill,' that has a more positive sentiment in a document describing video games than when used in a news article. Similarly, words used in social media and medical domains tend to be atypical and idiosyncratic as compared to generic language use.

DA algorithms and related transfer learning algorithms are often successfully used to perform sentiment analysis, where given a document, the sentiment label for the document is to be determined. Since adaptation can be viewed as a shift in the position of words within the embeddings spaces across the domains of interest, shifts in the spatial position of the words can be measured to verify that the algorithm is indeed capturing domain knowledge. Once it is established that a given DA method is capturing domain semantics at the word level, the DA word embeddings can then be used as input to an encoder+classification layer to perform sentiment analysis.

This paper applies a generic adaptation framework to tasks that make use of domain semantics. Contributions of this paper are fourfold. i) First, we propose a generic domain adaptation layer that can be interfaced with any neural network block with the aim of improving performance on a downstream sentiment classification task. ii) Second, we measure the significance of the shift in word position when represented in a generic embedding space and when represented in the adapted embedding space. The strategy is to construct DA word embeddings as in~\cite{P18-2007} and make use of a seed lexicon as in~\cite{P18-1075} and~\cite{mikolov2013exploiting}. However, rather than use the seed lexicon to transform words from two different domains into a common space, we use the seed lexicon to verify that the DA word embeddings have indeed captured significant domain semantics. iii) We test our hypothesis on a recently introduced~\cite{friedland2017civic}\footnote{\url{https://github.com/naacl18sublong/Friedland}} data set of tweets from Liberal and Conservative Twitter users from the state of Wisconsin. As illustrated in~\cite{P18-1228}, political discourse makes for an interesting setting to study domain adaptation. (iv) Via thorough experimentation we show that using a DA layer helps improve the performance of standard architectures on sentiment analysis tasks by $2-8\%$ on various binary and multi-class balanced and imbalanced data sets. We also, show that our DA architectures outperform sophisticated architectures such as BERT, LR-Bi-LSTM, Self-attention by $1-2\%$.

The rest of this paper is organized as follows, Section~\ref{relwork} discusses related work. Section~\ref{model} describes in detail, the proposed algorithm, Section~\ref{exps} presents the experimental results and Section~\ref{conc} concludes this work.

\section{Related Work}\label{relwork}
This paper discusses domain adaptation techniques and the applicability of existing algorithms to diverse downstream tasks. Central themes of this paper tie into word level domain semantics, while being related to the overall objective of domain adaptation.

Recent work such as SEMAXIS~\cite{P18-1228} investigates the use of word level domain semantics for applications beyond sentiment analysis. The authors introduce the concept of a semantic axis based on word antonym pairs to capture semantic differences across corpora. Similarly, work by~\cite{hamilton-EtAl:2016:EMNLP2016} captures domain semantics in the form of sentiment lexicons via graph propagation. While both these lexical based approaches are similar to the ideas of this paper, a major difference is that like~\cite{P18-2007}, we do not make use of any predefined domain specific lexicons to capture domain semantics. Our idea is to use word occurrences and contexts to provide a raw estimate of the domain semantics. Using generic embedding spaces as baselines, adaptation is performed by projecting generic embeddings into a learned `adaptation' space.

Typical downstream applications such as cross lingual and/or multi-domain sentiment classification, using algorithms proposed by~\cite{P18-1075},~\cite{N18-1050}, make use of DNNs with RNN blocks such as BiLSTMs to learn both generic and domain specific representations. Particularly, work focused on multi-domain sentiment classification as in~\cite{liu2016recurrent},~\cite{nam2016learning}, proposes building neural architectures for each domain of interest, in addition to shared representation layers across all domains. While these techniques are effective, they are not ideal in domains with limited data.

On the other hand work such as ELMo~\cite{N18-1202} and BERT~\cite{devlin2018bert} propose deeply connected layers to learn sentence embeddings by exploiting bi-directional contexts. While both methods have achieved tremendous success in producing word (ELMo) and sentence (BERT) level encodings that perform well in several disparate NLP tasks such as question-answer solving, paraphrasing, POS tagging, sentiment analysis etc, these models are computationally expensive and require large amounts of training data. Particularly when used in a transfer learning setting, both algorithms assume that a large amount of data is present in the source as well as the target domains. In contrast, our proposed adaptation layer is particularly well suited in applications with limited data in the target domain.

Our proposed algorithms depart from these approaches by capturing domain semantics through shallow layers for use with generic encoder architectures. Since some of the most successful algorithms in text classification~\cite{kim:2014:EMNLP2014} and sentence embeddings~\cite{conneau-EtAl:2017:EMNLP2017} make use of CNN and BiLSTM building blocks, we suggest a generic adaptation framework that can be interfaced with these standard neural network blocks to improve performance on downstream tasks, particularly on small sized data sets.

\section{Shallow Domain Adaptation}\label{model}
\begin{figure}
\centering
\includegraphics[width=\columnwidth]{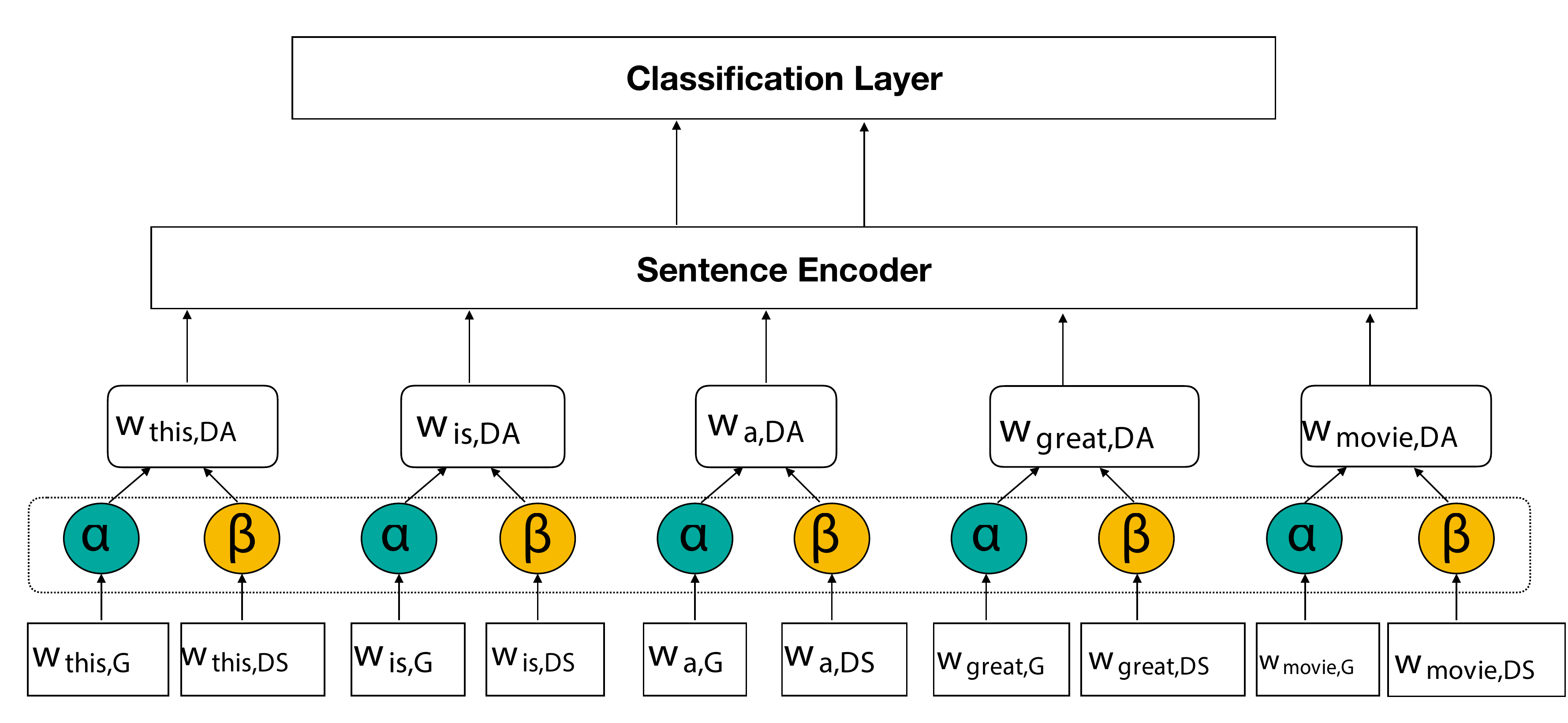}
\caption{This figure illustrates the three part neural network model. The first part is an adaptation layer. The second part is a generic sentence encoder and the last part is a classification layer. Inputs to the shallow adaptation layer are the generic and the domain specific (DS) word embeddings. Output of the adaptation layer is the domain adapted (DA) word embedding.}
\label{gen}
\end{figure}

This section introduces a `shallow' adaptation layer\footnote{Codes for shallow adaptation will be made available upon acceptance of paper for publication.}, to be interfaced with a neural network based sentence encoding layer, followed by a classification layer. Figure~\ref{gen} describes a generic framework of the proposed model. Brief descriptions of the three layers follow.

\noindent
\textbf{Adaptation Layer:} Generic word embeddings from GloVe~\cite{pennington-socher-manning:2014:EMNLP2014} or word2vec~\cite{mikolov2013distributed} are combined with an LSA-based \cite{deerwester1990indexing} Domain Specific (DS) word embedding using KCCA as proposed in~\cite{P18-2007}. Once the KCCA projections $(\bar{\vw}_{i,G}, \bar{\vw}_{i,DS})$ for the generic and DS word embeddings are obtained for a given word $i$, weights $\alpha$ and $\beta$ are learned such that the domain adapted word embedding is $\bar{\vw}_{i,DA} = \alpha \bar{\vw}_{i,G}+ \beta \bar{\vw}_{i,DS}$. Weights $\alpha$ and $\beta$ are learned by a single CNN layer. Since this layer learns only the weights used to obtain the DA embeddings, we call this a `shallow' adaptation layer.

The output of this layer (i.e., the DA embeddings), is used to initialize a generic sentence encoder. Output from the sentence encoder is sent to a classification layer and the classification error is back propagated to update weights of the adaptation layer. By keeping the sentence encoder fixed, (i.e., the parameters of the encoder are not updated during back propagation), this method can be generalized for use with any pre-trained sentence encoder. This is advantageous, particularly in data sets that are too small to be trained end-to-end. Another advantage is that adaptation is not performed on fully connected layers that need to be trained end-to-end. This considerably reduces the number of parameters that need to be learned.

\begin{figure*}
\centering
\includegraphics[width=\textwidth]{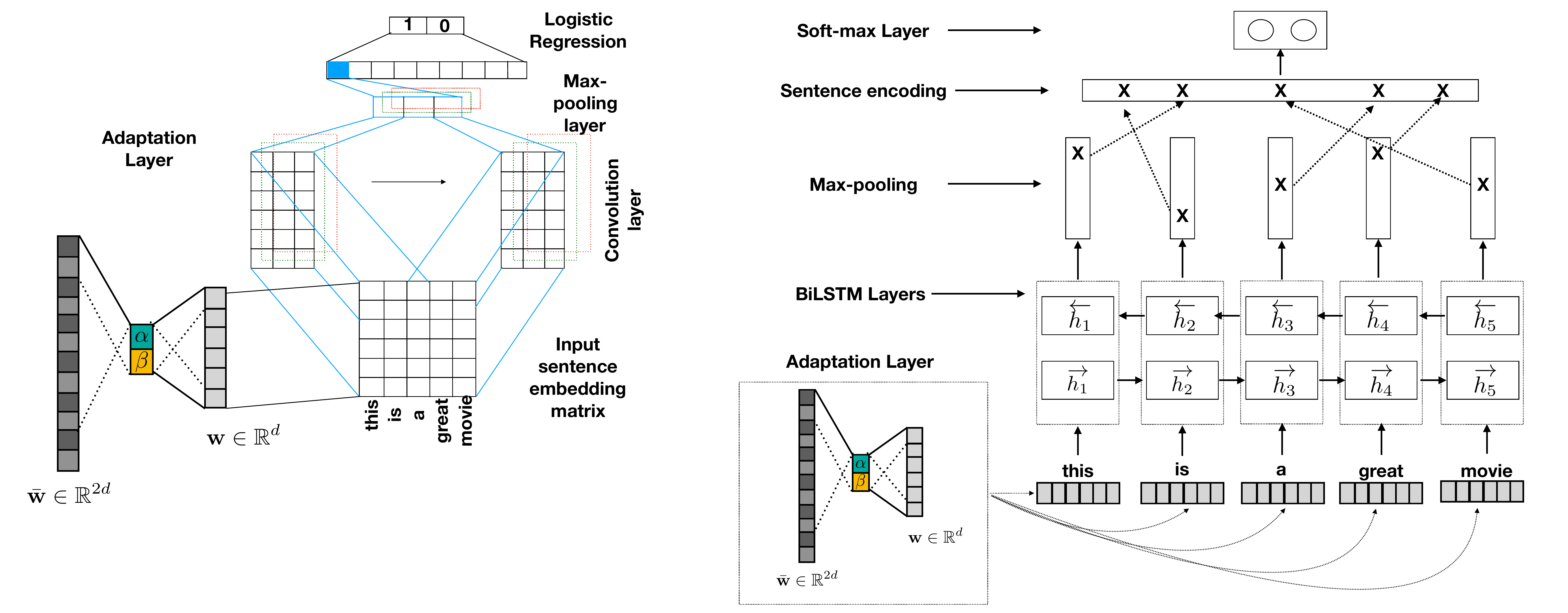}
\caption{\textbf{Left:} This figure illustrates the adaptation layer, that takes as input a $2d$ dimensional word embedding containing the generic and DS KCCA projections and learns a $d$ dimensional DA word embedding. This DA word embedding is then input to the CNN encoder. The entire network can be trained end-to-end to learn parameters $\alpha$ and $\beta$ in addition to the sentence embedding and classifier, or the network can be optimized to learn only the weights $\alpha$ and $\beta$ and weights of the classifier. \textbf{Right:} This figure illustrates the adaptation layer, that takes as input a $2d$ word embedding containing the generic and DS-KCCA projections and learns a $d$ dimensional DA word embedding. This DA word embedding is then input to the BiLSTM+max-pooling encoder. The entire network can be trained end-to-end to learn parameters $\alpha$ and $\beta$ in addition to the sentence embedding and classifier, or the network can be optimized to learn only the weights $\alpha$ and $\beta$ and weights of the classifier.}
\label{comb}
\end{figure*}

Figure~\ref{comb} illustrates the CNN encoder architecture and the Adaptation Layer. Input to the adaptation layer is a word embedding of dimension $2d$, where $d$ is the dimension of the generic and DS embeddings. For a given word, the generic and DS embeddings are concatenated and interleaved. A single layer CNN learns $\alpha$ and $\beta$ via a $2\times 1$ kernel. The domain adapted word $\vw_{i,DA}$ is then passed as input to the CNN encoder as shown in the figure. Note that we also use this framework with a BiLSTM sentence encoder as in~\cite{conneau-EtAl:2017:EMNLP2017} and is illustrated in Figure~\ref{comb}. Output of the BiLSTM units is max-pooled to obtain the sentence encoding.

\noindent
\textbf{Sentence Encoder:}
CNN encoder as in~\cite{kim:2014:EMNLP2014} and BiLSTM encoder proposed in~\cite{conneau-EtAl:2017:EMNLP2017} are used to encode sentences in this work. Since both models have been extensively discussed in literature we shall skip elaborating on the basic model architectures. We reproduced the basic encoder models for the experiments in Section~\ref{exps} with exception of few hyperparameters, since the choice of data sets in this work differs from that of~\cite{kim:2014:EMNLP2014} and~\cite{conneau-EtAl:2017:EMNLP2017}.

\noindent
\textbf{Classification Layer:} This layer learns weights for multi-class classification using a soft-max function.

\subsection{Evaluating KCCA for Domain Adaptation}
\label{motv}

In their work~\cite{P18-2007}, the authors demonstrate the effectiveness of KCCA based embeddings via classification experiments. However, in order to verify that the word level adaptation performed by the KCCA step can capture relevant domain semantics, we perform the following experiment.
\subsection{Experimental Setup}
Start with a dataset where language use is polarized. For example, the language used in the tweets of liberals likely differs from the language used in the tweets of conservatives as they express opinions on key political issues such as government, immigration, border control etc. These two vocabularies (Liberal and Conservative) are encoded in DS embeddings which are each paired with a generic embedding, and then again mapped into a common DA space. The words in this common space that are most different are then identified, and compared to a ground truth/gold standard list of keywords which represent ideas or concepts on which the two groups of users are known to have the most polarized opinions. Comparing the gold standard list with the list derived from the DA embedding demonstrates the efficacy of the method. This experiment adopts the list of politically contentious words that appear in \cite{li2017speaking}.

From the tokenized tweets of the users, construct a vocabulary of words $V_{Lib}$ and $V_{Con}$ that represent the Liberal and Conservative tweets respectively. Domain specific (DS) word embeddings are first constructed from $V_{Lib}$ and $V_{Con}$. Then, KCCA projections are obtained for generic GloVe embeddings and the DS word embeddings for each user group, in order to construct DA word embeddings for Liberal $\hat{\bw}_{DA(Lib)}$ and Conservative $\hat{\bw}_{DA(Con)}$ users.

Next consider the set of common words in the two vocabularies, i.e $V_{common} = V_{Lib} \cap V_{Con}$. Perform a second KCCA to bring DA words in $V_{common}$ from both Liberal $V_{Lib}$ and Conservative $V_{Con}$ vocabularies into a common subspace. To measure the `shift' $\psi$, calculate the $l_{2}$ distance for each word embedding $\hat{\bw}_{i} \in V_{common}$, i.e
\begin{align}
\psi=||\hat{\bw}_{i,DA(Lib)} - \hat{\bw}_{i,DA(Con)}||_{2} ~~~~ \forall i \in V_{common}
\end{align}

Words with large values of $\psi$ are considered to have shifted the most between the domains of liberal and conservative users. Words are then ordered based on the magnitude of $\psi$, and cross referenced with words in the gold standard list.

In order to demonstrate that the KCCA can capture domain semantics when learning DA word embeddings, the following section presents an analysis against a random baseline.
\subsection{Study of a random baseline}
A random baseline is one that picks $k$ words from $V_{common}$, randomly, and reports these $k$ words as having shifted the most with respect to word use within tweets that correspond to Liberal and Conservative users. This random baseline follows a hypergeometric distribution, which is a discrete probability distribution that calculates the probability of obtaining $k$ successes (of a sample with a specific feature) from $n$ draws, from a finite population of $V$, without replacement, where exactly $K$ samples contain the specific feature.

For example, suppose $|V_{common}|=V$, and say the number of words in the gold standard list are $K$. Then the probability of the random baseline picking $k < K$ words out of $n$ words is
\begin{align}
Pr(X=k) = \frac{{K \choose k}{V-K \choose n-k}}{{V \choose n}}.
\end{align}

To calculate these probabilities for the KCCA-DA word embeddings, we shall use the LibCon data set of Section
\ref{motv}. A gold standard list of 136 key topics/concepts important to Liberals and Conservatives is obtained from the study by~\cite{li2017speaking}, and additional details about the LibCon data set and the gold standard list can be found in Section~\ref{ds}.

From the LibCon data set $|V_{common}|=V=1573$. Out of these 1573 words, there are 74 words that are present in the gold standard list. Hence $K=74$. After obtaining the DA word embeddings for words in $V_{common}$ and calculating the shift $\psi$, we take the top 200 words that shifted most. From these words, at least 20 are present in the gold standard list.

The number of words-of-interest in a sample of $200$ words forms a hypergeometric distribution with mean $\mu = n\frac{K}{V}\approx 9.4$, and standard deviation 2.7984.

Direct calculation shows that the probability of picking exactly 20 words from this subset is $Pr(X=20) = 0.000346$.
Similarly, the probability that 20 or more would have been chosen by chance is,
\begin{align}
p = \sum_{k=20}^{200} Pr(X=k)
\end{align}
which, by numerical calculation, is $p=0.000524$. Such a result is highly unlikely to occur by chance. Accordingly, the KCCA domain adapted word embeddings are indeed capturing something significant about the language usage in the domain.

Table~\ref{words} presents words from the gold standard list with words in bold face indicating words that shift the most across Liberal and Conservative domains.

\begin{table}
\centering
\resizebox{0.5\textwidth}{!}{\begin{tabular}{l l l l l l l}
\hline
action&cost&dream&help&media&\textbf{reform}&\textbf{women}\\
\textbf{amendment}&court&education&honor&need&\textbf{republican}&work\\
\textbf{attack}&\textbf{crisis}&fact&hope&order&right&world\\
budget&deal&\textbf{force}&income&pledge&risk&\\
burden&debate&\textbf{freedom}&information&\textbf{police}&rule&\\
business&\textbf{debt}&fund&\textbf{insurance}&poll&school&\\
candidate&decision&funding&\textbf{justice}&power&spending&\\
care&defense&future&\textbf{labor}&president&\textbf{state}&\\
class&deficit&generation&leader&problem&\textbf{truth}&\\
college&\textbf{democrat}&\textbf{government}&leadership&program&value&\\
\textbf{congress}&development&\textbf{governor}&\textbf{legislature}&protection&\textbf{violence}&\\
\textbf{control}&divide&health&\textbf{legislation}&race&wealth&\\
\hline
\end{tabular}}
\caption{This table presents 74 words representing key political concepts common to Liberal and Conservative users on Twitter. Words in bold are the ones that `shift' the most (by $l_{2}$ distance) in use between Liberal and Conservative users on Twitter.}
\label{words}
\end{table}
\section{Experimental Results}\label{exps}
The discussion in Section~\ref{motv} provides motivation for adopting the KCCA projections as a tool to calculate DA word embeddings. These DA word embeddings can be combined with standard neural network models to obtain sentence embeddings that can be used to improve performance on downstream tasks such as sentiment analysis. To test the effectiveness of our proposed approach, we conduct a series of binary (LibCon, MR and SST) and multi-class (Beauty, Book and Music) classification tasks, using our adaptation layer on top of several standard architectures. The standard architectures used in this paper are architectures that are common in the current State-Of-The-Art (SOTA) for sentence encoders~\cite{conneau-EtAl:2017:EMNLP2017} and multi-domain sentiment analysis~\cite{N18-1050}. These architectures are a BiLSTM based architecture followed by some sort of pooling and then a soft-max classification layer.

Our main result is that using the proposed domain adaptation layer on top of existing architectures helps improve the performance of the architecture by about $2-8\%$. Furthermore, we show that basic architectures enhanced with our DA layer outperform SOTA architectures built specifically for domain adaptation and transfer learning problems. We now give a description of the various datasets that we use and the various baseline algorithm used for our experimental comparisons.

\subsection{Data Sets}\label{ds}
Motivated by the fact that our DA layer is most useful when the target dataset is of modest size, we present our experimental results on three such datasets. Furthermore, the text used in these datasets often uses the same set of words to express clearly contrastive sentiments, thereby setting up groups or domains of distinct word use within a single data set.

\noindent
\textbf{LibCon:} The LibCon (\textbf{Lib}eral and \textbf{Con}servative) data set is obtained from a study that aims towards understanding, analyzing and modeling the changing landscape of political discourse of an entire state over a duration of 10 years~\cite{friedland2017civic}. A key aspect of this study is the use of  Twitter data to analyze the latent network structure among key players of Wisconsin's political-media ecology. To do this, data was drawn from a collection of Twitter data housed at the UW-Madison School of Journalism and Mass Communication.
Data is drawn from a 10$\%$ sample of Twitter messages worldwide from Twitter's API (Twitter 2012). Note that the period of data collection corresponds to the re-election efforts of Governor Walker (of Wisconsin). Snowball sampling is employed to identify important Twitter handles active in four 28-day periods in the first half of 2012. A hand curated list of Twitter handles corresponding to key players in Wisconsin's political climate is identified. Combined with Twitter handles of nationwide users that interact frequently with users in this list, the final list of Twitter handles of interest is formed.

Twitter handles are confirmed to their corresponding political associations by verifying their party membership. The Twitter handle pool includes a balanced representation from Republican and Democratic accounts.
In depth description of data collection methodologies and outcomes can be found in~\cite{friedland2017civic} and readers are encouraged to consult this resource for additional information. A subsample of 1000 tweets, each from known Liberal (Democratic party) and Conservative (Republican party) users are considered for experiments here.

Gold standard list of topics key to both Democrat and Republican parties are obtained by analyzing primary debate data over a period of almost 20 years (1999-present)~\cite{li2017speaking}. A combination of modeling techniques involving semantic spaces and neural network architectures is used to arrive at the final list of 136 top/key concepts such as `Army', `Border', `Democracy', `Justice' etc. 
Out of these 136 words, 74 occur in the LibCon data set as seen in Table~\ref{words}.
\noindent

\textbf{Movie Review:} This is a benchmark dataset~\cite{pang2005seeing} from which we randomly sample 2500 positive and 2500 negative reviews for experiments.
\noindent

\textbf{SST:} The Stanford Sentiment Tree (SST) bank is yet another standard benchmark data set with reviews binarized to `positive' or `negative'. Our experiments use a 5000 sample training data set and a pre-determined test set of 5000 points.

\noindent

\textbf{Beauty, Book, Music:} These data sets each consist of 6000 beauty product, book and music reviews with a balanced distribution of three class labels. This data set was introduced by~\cite{D18-1383}. Data is obtained by sampling from a larger data set of product reviews obtained from Amazon~\cite{mcauley2015image}. Labels are `positive', `negative' and `neutral'. An imbalanced setting for each data set is also available, where roughly 80$\%$ of the data points are `positive' and the remaining are `negative' and `neutral'. Details about number of data points for each data set can be found in the supplement.

\subsection{Baselines}
For all test data sets, in addition to contrasting the performance of the proposed adaptation layer added to a CNN and BiLSTM encoder, we also present comparisons against `Vanilla' CNN and BiLSTM encoders. For popular data sets like the MR and SST data sets we present results against some additional baselines, all of which are variants of the basic vanilla RNN encoders. The baselines that we used in this paper are,

\noindent
\textbf{BoW:} In this standard baseline each sentence is expressed as a weighted sum of its constituent word embeddings. Weights used are raw word counts. Both generic (GloVe) and DA word embeddings are used in this baseline. Note that the DA embeddings used in this baseline are the same as in~\cite{P18-2007}, i.e $\alpha= \beta=0.5$

\noindent
\textbf{Vanilla CNN:} A CNN based sentence classification framework as introduced by~\cite{kim:2014:EMNLP2014} is used as a `Vanilla' CNN baseline.

\noindent
\textbf{Vanilla BiLSTM:} A Basic BiLSTM encoder followed by max pooling as proposed by~\cite{conneau-EtAl:2017:EMNLP2017} is used as the `Vanilla' BiLSTM baseline. Note that the vanilla baseline contains no additional features like attention.

\noindent
\textbf{BERT:} Bidirectional Encoder Representations from Transformers (BERT)~\cite{devlin2018bert} is a generic encoder framework that learns sentence embeddings by jointly conditioning on both left and right context in all layers. The Bert encoder is generic and requires a tuneable output layer to be learned in order to perform a chosen task. In our work, we use the pooled output from the Bert encoder to obtain sentence representations for all our data and then learn weights for a classifier to perform classification on the given data set.

\noindent
\textbf{Self-attention:} A model that generates interpretable, structured sentence embeddings using Self-attention mechanism~\cite{lin2017structured}.

\noindent
\textbf{LR-Bi-LSTM:} A model that imposes linguistic roles to Bi-LSTM architectures~\cite{P17-1154}.

\noindent
\textbf{DAS:} The Domain Adaptive Semi-supervised algorithm~\cite{D18-1383} is a transfer learning based solution that performs sentiment classification. In their experiments, the authors train on review data from different source domains such as book, music and electronics and test on a target domain such as beauty. In contrast, our method does not train on these three sources; rather, it uses only the Beauty reviews to learn the DS word embeddings (and then combines these with the generic pretrained embedding to learn the DA embedding used for classification. On the balanced data sets, metric reported is Accuracy, the same used by~\cite{D18-1383}. On the imbalanced data sets, in addition to accuracy we report micro f-score.

\subsection{Experimental methodology}
The LibCon, MR and Beauty, Book and Music data sets do not have dedicated train/dev/test splits, so we created $80/10/10$ splits. For the SST dataset we make use of pre-defined test data set of 2210 data points. When pre-tuned models are available, such as for BERT, we further fine-tune it using our train/dev splits. All the baselines use the same splits of data.

\begin{table*}
\centering
\resizebox{2\columnwidth}{!}{\begin{tabular}{|l|c|c|c|c|c|c|c|c|c|c|}
\hline
\multirow{2}{*}{Algorithm}&\multirow{2}{*}{Acc on LibCon}&\multirow{2}{*}{Acc on Beauty (B)}&\multirow{2}{*}{Acc on Book (B)}&\multirow{2}{*}{Acc on Music (B)}&\multicolumn{2}{|c|}{Beauty (I)}&\multicolumn{2}{|c|}{Book (I)}&\multicolumn{2}{|c|}{Music (I)}\\
\cline{6-11}
&&&&&Acc & F-score&Acc&F-score&Acc&F-score\\
\hline
BoW (GloVe)&64.0&56.18&69.1&66.5&74.8&74.8&66.08&66.08&75.8&75.8\\
BoW (DA embeddings, $\alpha=\beta=0.5$)&65.3&59.02&71.9&67.6&75.12&75.12&67.0&67.0&77.12&77.12\\
Vanilla CNN&61.36&61.7&66.5&61.5&74,5&74.6&76.8&78.1&77.5&77.5\\
Vanilla BiLSTM&68.5&75.6&77.6&80.0&84.3&84.3&83.4&83.4&84.2&84.2\\
Adapted CNN&63.24&62.8&71.0&63.5&80.1&75.0&78.0&76.6&78.1&78.1\\
\textbf{Adapted BiLSTM}&\textbf{72.1}&\textbf{77.3}&\textbf{80.3}&\textbf{81.5}&\textbf{85.2}&\textbf{85.2}&\textbf{84.7}&\textbf{84.7}&\textbf{85.6}&\textbf{85.6}\\
BERT&70.3&-&-&-&-&-&-&-&-&-\\
DAS&N/A&56.0&67.6&58.6&54.8&-&61.06&-&55.1&-\\
\hline
\end{tabular}}
\caption{This table presents performance (accuracy score) of the baseline algorithms on the LibCon and Beauty, Book and Music data sets in balanced and imbalanced setting. Micro F-score is reported on the imbalanced Beauty, Book and Music data sets as well. Bold face indicates best performing algorithm.}
\label{exps1}
\end{table*}


\textbf{Hyperparameters:} All word embeddings-GloVe, DS and KCCA projections used to obtain the DA embeddings are of dimension 300. Both the CNN and BiLSTM encoders learn sentence embeddings of 300 dimensions on all data sets, except on the LibCon data set where the CNN encoders have a filter size of 120. Dropout probability is set to 0.5. The rest of the hyperparameters used are the same as in~\cite{D18-1383} and are found in the supplement.
When using BERT, the `bert-base-uncased' model is used to obtain sentence embeddings. Pooled output from the BERT encoder is used to represent the input sentence. The classifier learned during tuning on the train data sets is similar to the linear classification layer introduced in the BERT classification framework, with the exception of an additional Tanh activation. The additional Tanh activation function was used in order to obtain comparable results for BERT and other baselines. The size of test data is particularly challenging for a large model like BERT that has several learnable parameters, even when used only for fine-tuning the pre-trained model.
The BERT classification framework is reproduced for use in our experiments, reflecting the original tuning scripts to the best of our ability. Additional information about hyperparameters can be found in the supplement.

\begin{table}[!h]
\resizebox{\columnwidth}{!}{\begin{tabular}{|l|l|l|}
\hline
Algorithm & Accuracy on MR& Accuracy on SST\\
\hline
BoW (Generic)&75.7*&48.9*\\
BoW (DA embeddings)&77.0*&49.2*\\
Vanilla CNN& 72.5*&49.06*\\
Vanilla BiLSTM& 81.8*&50.3*\\
LR-Bi-LSTM&82.1&50.6\\
Self-attention&81.7&48.9\\
Adapted CNN&80.8*&50.0*\\
\textbf{Adapted BiLSTM}&\textbf{83.1*}&51.2\\
BERT&74.4*&\textbf{51.5}\\
\hline
\end{tabular}}
\caption{This table presents results for the MR and SST data sets. Reported performance metric is accuracy. Results with * indicate that performance metric is reported on the test dataset after training on a subset of the original data set. Bold face indicates best performing algorithm}
\label{exps2}
\end{table}

\begin{table}[!h]
\resizebox{\columnwidth}{!}{\begin{tabular}{|l|l|l|}
\hline
Algorithm & Accuracy on MR& Accuracy on SST\\
\hline
Vanilla CNN (1000 pts)&65.1&48.8\\
Adapted CNN (1000 pts)&74.7&49.6\\
Vanilla BiLSTM(1000 pts)&76.9&50.08\\
Adapted BiLSTM (1000 pts)&78.1&50.3\\
\hline
Vanilla CNN (2500 pts)&66.5&49.0\\
Adapted CNN (2500 pts)&77.4&50.7\\
Vanilla BiLSTM(2500 pts)&78.8&50.2\\
Adapted BiLSTM (2500 pts)&80.1&51.0\\
\hline
\end{tabular}}
\caption{This table presents the accuracy obtained by Vanilla and adapted baselines on smaller subsamples of the training data for the MR and SST data sets.}
\label{exps3}
\end{table}

\subsection{Results}\label{results}
Table~\ref{exps1} presents results on the LibCon and the balanced (B) and imbalanced (I) Beauty, Book and Music data sets and Table~\ref{exps2} presents results on the SST and MR data sets. The performance metric reported in both tables is accuracy with with additional micro f-scores reported in Table~\ref{exps1}.

From Table~\ref{exps1}, it is observed that on the LibCon data set, where we have a considerable difference in language use between the two groups of users, the adapted BiLSTM and adapted CNN perform much better than the vanilla baselines. Furthermore, our proposed adaptation layer improves the performance of the Vanilla BiLSTM to surpass the performance of fine-tuned BERT. Note that the BERT encoder is pre-trained on all of Wikipedia, a much larger training data set than used for training the Adapted BiLSTM encoder. The adapted BiLSTM and CNN encoders also outperform DAS on the multiclass Beauty, Book and Music data sets. Performance of LSTMs for text surpass that of CNNs as seen in current literature. This observation is consistent in our results as well. Since the DAS baseline makes use of a CNN encoder, it is expected that a very basic LSTM such as the Vanilla BiLSTM would perform much better than any CNN framework.
Results Table~\ref{exps3} strengthens our hypothesis that the adaptation layer is particularly well suited for small training and test data regimes.

Note that on the MR data set, the performance of fine-tuned BERT is lower than the BiLSTM baselines. While the size of training data used to optimize BERT, makes the encoder an excellent generic sentence encoder, the small size of the tuning data sets poses a considerable challenge. This is because, when the BERT encoder is fine tuned for a particular task, output from the encoder is treated as a fixed feature on top of which it learns a task specific layer. Using pre-trained BERT provides a high quality feature vector that can be used for applications on generic data sets, but this does not assure good performance when tuned on a small data set where words may be used idiosyncratically. On the other hand, on a highly polarized data set like the LibCon data set, while the distinct linguistic features enable the BERT encoder to capture good sentence embeddings, the size of the data set still poses a challenge.

\noindent
\textbf{Varying Size of Training Data:} To further illustrate the effectiveness of our method on data sets with limited training data, we train the Vanilla and adapted BiLSTM and CNN encoders with even smaller samples of the training data. Our results are consistent on varying size of data sets and performance metrics for the same can be found in the supplement. Table~\ref{exps3} compares accuracy on the MR and SST data sets, of Vanilla and adapted encoders trained with 1000 and 2500 points. Repeating this experiment with BERT resulted in tremendous over-fitting on the test set and so we do not present results from BERT in Table~\ref{exps3}.

\noindent
\textbf{Qualitative analysis of DA word embeddings:} Here we present a small analysis of the DA embeddings obtained from the weights learned via the shallow adaptation layer. The analysis presented here is on the Book review data set consisting of roughly 10000 unique word tokens. Weights learned via the CNN encoder on this data set are $\alpha = 0.7145$ and $\beta = 0.3994$. To compare the DA embeddings against the GloVe common crawl embeddings, we first standardize both sets of embeddings to have an average norm of 1. Then we compute shift via the $l_{2}$ distance as described in Section~\ref{motv}. Out of 100 most shifted words, Adjectives (21$\%$) and Adverbs (22$\%$) such as `emotional', `profound' and `emotionally' and `obviously' shift the most. These words are used extensively when writing reviews about books. On the other hand words that shift the least are Nouns (69$\%$) such as `Higgins', `Gardner' and `Schaffer' that correspond to author and character names. This observation is consistent with the domain adaptation hypothesis. A data set of book reviews will use Adjectives and Adverbs differently in-domain as opposed to out-of-domain. On the other hand, Nouns and Noun Phrases such as character and author names remain fixed regardless.

\section{Conclusions and Future Work}
\label{conc}
This paper shows that domain semantics captured in adapted word embeddings can improve the performance of (pre-trained) encoders in downstream tasks such as sentiment classification; particularly on data sets where obtaining data sufficient enough for tuning pre-trained encoders or for end-to-end training is difficult. Experiments show the effectiveness of the method in binary and multiclass classification tasks. The proposed framework outperforms competing transfer learning based algorithms in overcoming limitations posed by the size of training data, while learning fewer parameters than an end-to-end trained network along with obtaining better results on classification tasks. 
Recent work that focuses on capturing domain semantics in word embeddings, demonstrate the effectiveness of such approaches mostly via performance on a downstream task such as sentiment analysis/classification alone. 
In Section~\ref{motv} we provide a simple yet effective demonstration of how techniques such as KCCA, that are used to capture domain semantics, can capture relevant semantics in the learned word embeddings.
As future work, we will interface the adaptation layer for use recently introduced large scale language models such as BERT. Due to size and memory complexities of available implementation of large scale language models such as BERT, an easy integration with our proposed adaptation framework does not currently exist.




\bibliography{refbib}
\bibliographystyle{acl_natbib}
\end{document}